\documentclass[journal,twoside,web]{ieeecolor}
\usepackage{jsen}
\usepackage{cite}
\usepackage{amsmath,amssymb,amsfonts}
\usepackage{algorithmic}
\usepackage{graphicx}
\usepackage{textcomp}
\usepackage{wrapfig}
\usepackage{booktabs}
\usepackage{here}
\usepackage{amsmath,amssymb,amsfonts}
\def\BibTeX{{\rm B\kern-.05em{\sc i\kern-.025em b}\kern-.08em
    T\kern-.1667em\lower.7ex\hbox{E}\kern-.125emX}}
\definecolor{abstractbg}{rgb}{0.89804,0.94510,0.83137}
\begin{document}
\title{Data-driven sparse sensor placement based on A-optimal design of experiment with ADMM}
\author{Takayuki Nagata, Taku Nonomura, Kumi Nakai, Keigo Yamada, Yuji Saito, Shunsuke Ono, \IEEEmembership{Member, IEEE}
\thanks{This article was submitted on September xx. This work was supported by JST CREST Grant Number JPMJCR1763, Japan.}
\thanks{T. Nagata is with the Department of Aerospace Engineering, Tohoku University, Miyagi, JAPAN (nagata@aero.mech.tohoku.ac.jp). }
\thanks{T. Nonomura is with the Department of Aerospace Engineering, Tohoku University, Miyagi, JAPAN (nonomura@aero.mech.tohoku.ac.jp).}
\thanks{K. Nakai is with the Department of Aerospace Engineering, Tohoku University, Miyagi, JAPAN (nakai@aero.mech.tohoku.ac.jp).}
\thanks{K. Yamada is with the Department of Aerospace Engineering, Tohoku University, Miyagi, JAPAN (yamad.keigo@aero.mech.tohoku.ac.jp).}
\thanks{Y. Saito is with the Department of Aerospace Engineering, Tohoku University, Miyagi, JAPAN (saito@aero.mech.tohoku.ac.jp).}
\thanks{S. Ono is with the Department of Computer Science, Tokyo Institute of Technology, Tokyo, JAPAN (ono@isl.titech.ac.jp).}}

\IEEEtitleabstractindextext{%
\fcolorbox{abstractbg}{abstractbg}{%
\begin{minipage}{\textwidth}%
\begin{wrapfigure}[12]{r}{3in}%
\includegraphics[width=3in]{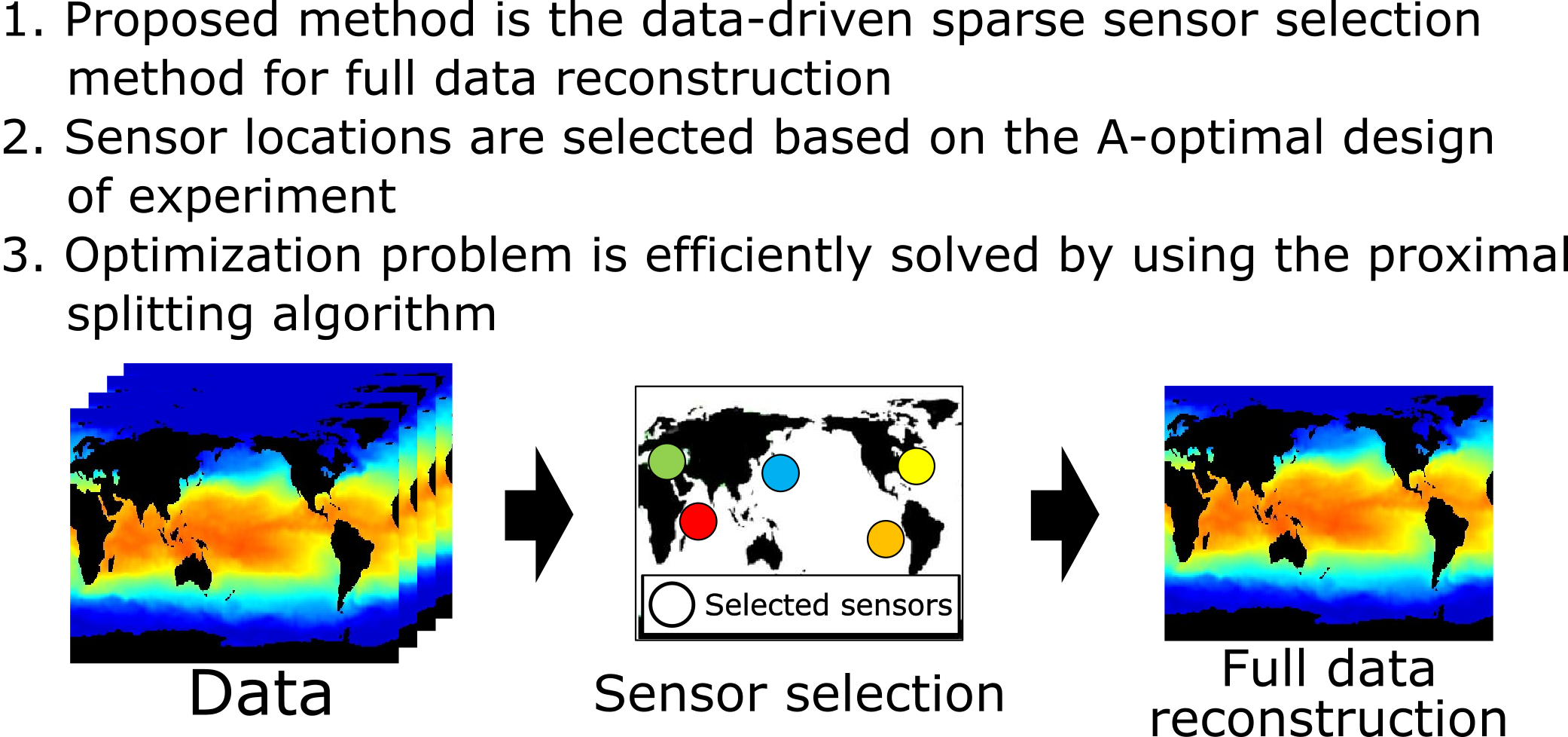}
\end{wrapfigure}%
\begin{abstract}
The present study proposes a sensor selection method based on the proximal splitting algorithm and the A-optimal design of experiment using the alternating direction method of multipliers (ADMM) algorithm. The performance of the proposed method was evaluated with a random sensor problem and compared with the previously proposed methods such as the greedy method and the convex relaxation. The performance of the proposed method is better than an existing method in terms of the A-optimality criterion. In addition, the proposed method requires longer computational time than the greedy method but it is quite shorter than the convex relaxation in large-scale problems. The proposed method was applied to the data-driven sparse-sensor-selection problem. A data set adopted is the NOAA OISST V2 mean sea surface temperature set. At the number of sensors larger than that of the latent state variables, the proposed method showed similar and better performances compared with previously proposed methods in terms of the A-optimality criterion and reconstruction error.
\end{abstract}

\begin{IEEEkeywords}
Alternating direction method of multipliers, optimal design of experiment, sensor selection, sparse observation
%Enter key words or phrases in alphabetical 
%order, separated by commas. For a list of suggested keywords, send a blank 
%e-mail to keywords@ieee.org or visit \underline
%{http://www.ieee.org/organizations/pubs/ani\_prod/keywrd98.txt}
\end{IEEEkeywords}
\end{minipage}}}

\maketitle

\section{Introduction}
\label{sec:introduction}
\IEEEPARstart{O}{bservation} of data and analysis, or control using observed data, is an important topic in various fields. Observation may be performed by surface or volume measurements, but generally, it is performed by discretely installed sensors. There is a constraint on the physical domain and on the cost of the infrastructure related to sensors. Further, when the measurement target is a structure such as robots or vehicles, additional restrictions appear depending on the size and structure of the measurement target. In such cases, it is necessary to maximize the information obtained with as few sensors as possible.

The global positioning system requires to select the optimal subset of the visible satellites for minimizing the geometric dilution of precision \cite{kihara1984satellite,phatak2001recursive}. This is the sensor selection problem selecting the optimum (in terms of the quality of acquired data) $p$ sensors from $n$-potential sensors locations, and such kind of sensor selection problem can be seen in various types of measurements such as acoustic measurements \cite{douganccay2009optimal,macho2016optimal}, structural health monitoring \cite{worden2001optimal,yi2011optimal}, and environment monitoring \cite{du2014optimal}.
%\cite{worden2001optimal,meo2005optimal,yi2011optimal}, and environment monitoring \cite{du2014optimal}.

The position of sensors is important in terms not only of the quality of observation but also of the cost of the sensor systems. In the case of wireless sensor networks, not only should the sensors be informative, but they should also communicate efficiently for energy conservation and a longer lifetime of the system \cite{krause2006near,masazade2012sparsity}.
%\cite{cardei2005maximum,krause2006near,ferentinos2007adaptive,masazade2012sparsity}.

Furthermore, Manohar et al. \cite{manohar2018data} proposed a framework for full state reconstruction based on sparse observation and a tailored library of features extracted from training data. Their framework takes advantage of the low dimensionality in high-dimensional data and realizes the compressed sensing using low-dimensional representation extracted by a modal decomposition such as the proper orthogonal decomposition \cite{berkooz1993proper}.

There are several approaches to solve sensor selection problems. Table~\ref{tab:mothods} shows the characteristics of the algorithms typically used in a sensor selection problem. The exact solution of the sensor selection problem can be obtained using global optimization techniques such as branch and bound \cite{welch1982branch,lawler1966branch}, but it can only be used in the problem of choosing a small number of sensors from a small number of potential sensors because of its expensive computational cost. Therefore, an approximated method based on the greedy method and convex optimization method has been studied.

%greedy QR
The greedy method can obtain the solution of a sensor selection problem with a small computational cost, although the obtained solution is a local optimal solution and the objective function is less flexible. 
Manohar et al. \cite{manohar2018data} proposed the QR-based greedy method which is related to the discrete empirical interpolation method \cite{chaturantabut2010nonlinear} and the QR-based discrete empirical interpolation method \cite{drmac2016new} in the framework of the Galerkin projection \cite{rowley2004model}. In their method, the sparse sensors discovered using singular value decomposition and QR pivoting. 
Saito et al. \cite{saito2019determinantbased} proposed the determinant-based greedy method. Their method selects sensors based on the D-optimal design of experiment. Their formulation is mathematically the same as the QR-based method for the underdetermined system, and the performance of the selected sensors in the overdetermined system is better than that of the QR-based method. In addition, the computation is accelerated by both the determinant formula and the matrix inversion lemma. Also, they extended the QR-based greedy method for vector-measurement sensor problems where the measured data has multiple components \cite{saito2020data}. Yamada et al. \cite{yamada2019fast} proposed the noise-robust greedy method by extending the determinant-based greedy method. The sensors are selected by maximizing the determinant of the matrix which corresponds to the inverse matrix appearing in the Bayesian estimation operator. Their method includes the covariance noise matrix in the objective function and can select the optimal sensor location while considering the correlated noise.
Nakai et al. \cite{nakai2020effect} adopted the E- and A-optimal design of experiment onto the sensor selection using the greedy method and investigated the effect of the objective function on the performance of the sensor selection algorithm using the greedy method. Their results suggested that the greedy method based on the D- and A- optimality criteria show a similar performance.

%
%DG
%Saito et al. \cite{saito2019determinantbased} proposed the determinant-based greedy sensor selection algorithm. Their method selects sensors based on the D-optimal design of experiment and is mathematically the same as the QR method for the number of sensors is less than or equal to the state variables. The computation is accelerated by both determinant formula and matrix inversion lemma than the QR-based greedy method. They also extended the QR-based greedy method for vector-measurement sensor problems where the measured data has multiple components \cite{saito2020data}.
%Yamada et al. \cite{yamada2019fast} proposed the noise-robust greedy sparse-sensor-selection method by extending the determinant-based greedy method. The sensors are selected by maximizing the determinant of the matrix which corresponds to the inverse matrix appearing in the Bayesian estimation operator. %The Bayesian estimation of state variables using the optimized sensors and the proir information robustly works even in the presence of the correlated noises.

%proximal methodを使った凸最適化によるセンサ選択
On the other hand, a method based on the convex optimization is also used to solve the sensor selection problem. Joshi and Boyd \cite{joshi2009sensor} proposed an optimization methodology for the convex relaxation problem of the sensor selection. Unlike the greedy method, their method can obtain a global optimal solution of the relaxed problem. The computational complexity of their method is cubic order of the problem size, and thus, the computational cost is quite smaller than that of the global optimization technique but still large to handle large-scale problems, though the  acceleration method using randomized algorithm was recently proposed.\cite{nonomura2020randomized} Furthermore, there is another method which is the convex optimization based on proximal splitting algorithms. The sparsity-promoting framework based on the proximal splitting algorithm was introduced by Fardad, Lin, and Jovanovi\'{c} \cite{fardad2011sparsity} and Lin, Fardad, and Jovanovi\'{c} \cite{lin2013design}. Their methods can obtain block sparse feedback and observer gains as well as select actuators and sensors.
Their framework was extended by Dhingra, Jovanovi\'{c} and Luo \cite{dhingra2014admm}, and they developed a customized algorithm using the alternating direction method of multipliers (ADMM) \cite{gabay1976dual,eckstein1992douglas} for a large-scale sensors and actuators selection problem in a dynamical system. %Also, Zare and Jovanovi\'{c} \cite{zare2018optimal} proposed the optimal sensor selection method via proximal optimization algorithm using the forward-backward quasi-Newton method. 
The sparsity-promoting framework allows us to exploit the separability of the objective function and the sparsity promoting term and to split the optimization problems into subproblems that can be solved analytically, and thus, corresponding optimization can efficiently be solved. 
The sensor selection methods based on proximal splitting algorithms have been applied to complex problems such as sensor selection for dynamical systems, but the sensor selection method for nondynamical systems based on the proximal splitting algorithm has not been proposed. There is a possibility to construct an efficient sensor selection algorithm in a wide range of problems by proposing a sensor selection method for nondynamical systems based on proximal splitting algorithms.
%\cite{gabay1976dual,eckstein1992douglas,boyd2011distributed}
%present study

In the present study, we consider the problem of selecting $p$ sensors, from among $n$-potential sensors. Each sensor gives an observation vector $\mathbf{y}$ of a linear function of latent variables $\mathbf{z}$ superimposed with independent identically distributed zero-mean Gaussian random noise. We propose a new method for sparse sensor selection problems based on proximal splitting algorithms (ADMM in the present study) and A-optimal design of experiment, and our goal is to choose the optimal sensor subset based on the optimal design of experiment to minimize the average error in the estimation. The contribution of the present paper is the following.
\begin{itemize}
    \item The sensor selection method for nondynamical systems based on proximal splitting algorithms was proposed, and several kinds of types of the sparsity promoting term were tested (the problem will be nonconvex depending on a sparsity-promoting term).
    \item The sensor selection in a large-scale problem can be solved with feasible computational cost (the computational complexity of the proposed method is the first order of the number of sensor candidates and the square order of the number of latent state variables) by the proposed method.
    \item The flexibility of the objective function is maintained like the convex relaxation.
\end{itemize}
%(computational complexity of the proposed method is the first order of the number of sensor candidates and the square order of the number of latent state variables)
%The computational cost is quite smaller than the convex relaxation thus 
%, and the performance of the proposed method is investigated in the random sensor problem. Our formulation can flexibly set the objective function like the convex relaxation method, and the computational cost is quite smaller than the convex relaxation method.

% Please add the following required packages to your document preamble:
% \usepackage{booktabs}
%\begin{table}[]
%\centering
%\caption{comparison of the characteristics of algorithms used in sensor selection.}
%\begin{tabular}{@{}l|lll@{}}
%\toprule
% & Exhaustive search & Convex relaxation & Greedy method \\ \midrule
%Solution & Exact solution & \begin{tabular}[c]{@{}l@{}}Global optimal\\ solution\end{tabular} & %\begin{tabular}[c]{@{}l@{}}Local\\ optimal solution\end{tabular} \\
%Cost & Quite expensive & Expensive & Small \\
%Objective function & - & Flexible & Not so flexible \\ \bottomrule
%\end{tabular}
%\label{tab:mothods}
%\end{table}
\begin{table}[]
\centering
\caption{comparison of the characteristics of algorithms used in sensor selection.}
\begin{tabular}{@{}l|lll@{}}
\toprule
 & Global optimization & Convex relaxation & Greedy method \\ \midrule
Solution & Exact & Global optimal & Local optimal \\
Cost & Quite expensive & Expensive & Small \\
Objective function & - & Flexible & Not flexible \\ \bottomrule
\end{tabular}
\label{tab:mothods}
\end{table}

\section{Algorithms}
\subsection{Formulation of Sensor Selection Problem}
A snapshot measurement of data $\mathbf{x}\in\mathbb{R}^n$ through sparse sensors can be explained as follows:
\begin{align}
    \mathbf{y}&=\mathbf{H}\mathbf{x} \\
            &\approx\mathbf{H}\mathbf{U}\mathbf{z} \\
            &\equiv\mathbf{C}\mathbf{z}
    \label{eq:observation}.
\end{align}
where $\mathbf{y}\in\mathbb{R}^p$, $\mathbf{z}\in\mathbb{R}^r$, and $\mathbf{C}\in\mathbb{R}^{p\times r}$ are the observation vector, the latent state vector, and the measurement matrix, respectively. The system above represents the problem to choose $p$ observations out of $r$ state variables by the sensors. The observation matrix $\mathbf{C}$ is the product of the sensor matrix $\mathbf{H}\in\mathbb{R}^{r\times n}$ and the sensor candidate matrix $\mathbf{U}\in\mathbb{R}^{n\times r}$. The sensor matrix $\mathbf{H}$ is a sparse matrix that indicates sensor locations. Each row vector of $\mathbf{H}$ is a unit vector and the position of unity in the row vectors indicates which point among $n$ sensor candidates is chosen. The sensor candidate matrix $\mathbf{U}$ is usually a tall-skinny matrix (i.e. $n>>r$) in a practical problem because $n$ corresponds to the degree of freedom in the spatial direction of spatiotemporal measurement data.
%センサを通した観測, Uはセンサ候補行列．モードが有るデータを見るなら低ランク近似

The estimated parameters $\mathbf{\tilde{z}}$ can be obtained by the pseudo-inverse operation.
\begin{align}
    \mathbf{\tilde{z}}
    =\mathbf{C}^{+}\mathbf{y}
    =\left(\mathbf{C}^{\mathsf{T}}\mathbf{C}\right)^{-1}\mathbf{C}^{\mathsf{T}}\mathbf{y}~~~~~(p>r) %\end{array} 
    \label{eq:LS_estimation}.
%left\{\begin{array}{cc}\bm{C}^{\mathrm{T}}\left(\bm{C}\bm{C}^{\mathrm{T}}\right)^{-1}\bm{y}, & p\le r \\   
\end{align}
Here, the notation of tilde above any variables or matrices indicates estimated parameters. Although it should be necessary to make a distinction between $p\leq r$ and $p>r$, we omit the formulation for $p<r$ because this study is concerned with $p>r$ only.

\subsection{A-optimality Criterion}  %copied from nakai-san's article
An optimality criterion provides a quality of measurement by computing the certain values of Fisher information matrix (FIM).
The A-optimal design of experiment minimizes the mean square error in estimating the parameter $\mathbf{\tilde{z}}$. Here, we consider the uniform independent Gaussian noises $\mathbf{v}=\mathcal{N}(0,\sigma^2)$ are imposed on the observation of data $\mathbf{x}$.
\begin{align}
    \mathbf{\tilde{z}}&=\left(\mathbf{C}^{\mathsf{T}}\mathbf{C}\right)^{-1}\mathbf{C}^{\mathsf{T}}\left(\mathbf{y}+\mathbf{v}\right) \nonumber \\
                &=\left(\mathbf{C}^{\mathsf{T}}\mathbf{C}\right)^{-1}\mathbf{C}^{\mathsf{T}}\left(\mathbf{H}\mathbf{x}+\mathbf{v}\right) \nonumber \\
                &=\left(\mathbf{C}^{\mathsf{T}}\mathbf{C}\right)^{-1}\mathbf{C}^{\mathsf{T}}\mathbf{C}\mathbf{z}+\left(\mathbf{C}^{\mathsf{T}}\mathbf{C}\right)^{-1}\mathbf{C}^{\mathsf{T}}\mathbf{v}
    \label{eq:ztilde}
\end{align}
Therefore, the error in estimation is the following.
\begin{align}
    \mathbf{\tilde z-z}&=\left(\mathbf{C}^{\mathsf T}\mathbf{C}\right)^{-1}\mathbf{C}^{\mathsf T}\mathbf{v}
   % \label{eq:esterr}
\end{align}
To consider the mean-variance in the estimation, the trace norm of the error covariance matrix is calculated.
\begin{align}
    &\mathrm{tr}\left(\mathrm{E}\left\{\left(\mathbf{\tilde z}-\mathbf{z}\right)\left(\mathbf{\tilde z}-\mathbf{z}\right)^{\mathsf T}\right\}\right) \\
    =&\mathrm{tr}\left(\mathrm{E}\left\{\left(\mathbf{C}^{\mathsf T}\mathbf{C}\right)^{-1}\mathbf{C}^{\mathsf T}\mathbf{vv}^{\mathsf T}\mathbf{C}\left(\mathbf{C}^{\mathsf T}\mathbf{C}\right)^{-1}\right\}\right)
\label{eq:esterr}
\end{align}
Here, the noise can be written by $\mathrm{E}\left(\mathbf{vv}^{\mathsf{T}}\right)=\sigma^2\mathbf{I}$ because of the Gaussian noises. Therefore,
\begin{align}
    \mathrm{E}\left(\mathbf{vv}^{\mathsf T}\right)=\sigma^2\mathbf{I},
    \label{eq:noiseatsens}
\end{align}
and Eq.~\ref{eq:esterr} becomes follows:
\begin{align}
    \mathrm{tr}\left(\sigma^2\left(\mathbf{C}^{\mathsf{T}}\mathbf{C}\right)^{-1}\right).
    \label{eq:trinv}
\end{align}

Hence, the objective function is the trace of the error covariance matrix in the A-optimal design of experiment. The sensor selection problem based on the A-optimality criterion can be expressed as the optimization problem 
\begin{align}
%    &\mathrm{minimize}\,\,f_{\mathrm{A}}\\
%    &f_{\mathrm{A}}\,=\,\left\{\begin{array}{cc}
%        \mathrm{tr}\,\left[\left(\bm{C}\bm{C}^{\mathrm{T}}\right)^{-1}\right], & p\le r \\
%        \mathrm{tr}\,\left[\left(\bm{C}^{\mathrm{T}}\bm{C}\right)^{-1}\right], & p>r
%    \end{array}\right.
\mathrm{minimize}\,\,\mathrm{tr}\,\left[\left(\mathbf{C}^{\mathsf{T}}\mathbf{C}\right)^{-1}\right].~~~~~(p>r)
\label{eq:obj_tr}
\end{align}
%To reduce the computational cost (quicken the computation), the matrix lemma is adopted/considered. Consequently, we can obtain the objective functions as follows:
% add derivation, proof
%\begin{align}
%    f_{\mathrm{A}}\,\approx\,\left\{\begin{array}{cc}
%        \quad\frac
%        {\bm{u}_{i_k}\bm{C}_{k-1}^{\mathrm{T}}\left(\bm{C}_{k-1}\bm{C}_{k-1}^{\mathrm{T}}\right) %^{-2}\;
%        \bm{C}_{k-1}\bm{u}_{i_k}^{\mathrm{T}} + 1}
%        {\bm{u}_{i_k}\left(\bm{I}-\bm{C}_{k-1}^{\mathrm{T}}\left(\bm{C}_{k-1}\bm{C}_{k-1}^{\mathrm%{T}}\right)^{-1}\bm{C}_{k-1}\right)\bm{u}_{i_k}^{\mathrm{T}}},
%        & p\le r \\
%        -\frac
%        {\bm{u}_{i}
%        \left(\bm{C}_{k-1}^{\mathrm{T}}\bm{C}_{k-1}\right)^{-2}
%        \bm{u}_{i}^{\mathrm{T}}}
%        {1+\bm{u}_{i}\left(\bm{C}_{k-1}^{\mathrm{T}}\bm{C}_{k-1}\right)^{-1}\bm{u}_{i}^{\mathrm{T}%}},
%        & p>r
%    \end{array}\right.\label{eq:obj_tr}
%\end{align}

\subsection{Problem Formulation}
%In the previous sensor selection algorithm, the sensor matrix $H$ (sparse sensor location) is directly acquired by the convex approximation and the greedy method. On the other hand, the sensor location is acquired 
Consider the gain matrix $\mathbf{K}\in\mathbb{R}^{p\times n}$ which recovers $\mathbf{z}$ from the observation $\mathbf{y}$ in Eq.~\ref{eq:observation},
\begin{align}
    \tilde{\mathbf{z}}&=\mathbf{Ky}\\ 
                      &=\mathbf{KUz}. 
\end{align}
Here, we assume that $\tilde{\mathbf{z}}$ is a unbiased approximation of $\mathbf{z}$. In this case,
\begin{align}
    \mathbf{KU}&=\mathbf{I}
\end{align}
should be satisfied. The average error in $\tilde{\mathbf{z}}$ can be explained in the same way described from Eq.~\ref{eq:ztilde}.
\begin{align}
    &\mathrm{tr}\left(\mathrm{E}\left\{\left(\mathbf{\tilde z}-\mathbf{z}\right)\left(\mathbf{\tilde z}-\mathbf{z}\right)^{\mathsf T}\right\}\right) \\
   =&\mathrm{tr}\left(\mathrm{E}\left[\mathbf{Kvv}^{\mathsf T}\mathbf{K}^{\mathsf T}\right] \right)=\mathrm{tr}\left(\sigma \mathbf{KK}^{\mathsf T} \right)
\label{eq:esterrD}
\end{align}
Here, $\tilde{\mathbf{z}}$ in the present formulation is follows:
\begin{align}
    \mathbf{\tilde{z}}&=\mathbf{KUz}+\mathbf{Kv}.
\end{align}
The gain matrix $\mathbf{K}$ should be the sparse matrix that has several nonzero column vectors to determine the sparse sensor location. Hence, drawing on the group-sparsity paradigm \cite{yuan2006model}, we augmented Eq.~\ref{eq:esterrD} with a sparsity-promoting term on the rows of $\mathbf{K}$. The group $L_0$-norm penalty is employed as a sparsity-promoting term as follows:
\begin{align}
    &\mathrm{minimize}~\mathrm{tr}\left(\mathbf{KK}^{\mathsf T}\right)
    +\lambda \left|\left|\left(\left|\left|\mathbf{k}_1\right|\right|_2,\cdots,\left|\left|\mathbf{k}_n\right|\right|_2\right)\right|\right|_0 \nonumber \\
    &~~~~\mathrm{subject~to}~\mathbf{KU=I},
    \label{eq:objorgGL0}
\end{align}
where ${\mathbf k}_i\in\mathbb{R}^n$ is the $i$-th column vector of $\mathbf{K}$. The sparsity-promoting term penalizes the number of nonzero columns of the $\mathbf{K}$ matrix, which correspond to the number of selected sensors.
As relaxed regularization of the group $L_0$-norm penalty, the group $L_1$-norm penalty is used. The objective function with group $L_1$-norm penalty becomes following.
%%%%%%%%%
\begin{align}
    \mathrm{minimize}~\mathrm{tr}\left(\mathbf{KK}^{\mathsf T}\right)+\lambda \sum_{i=1}^n\left|\left|\mathbf{k}_i\right|\right|_2~~~~\mathrm{subject~to}~\mathbf{KU=I}
    \label{eq:objorgGL1}
\end{align}
Furthermore, the other $L_0$-norm penalty, which value of the group $L_0$ norm is constrained, is introduced for a sensor selection problem.
\begin{align}
    &\mathrm{minimize}~\mathrm{tr}\left(\mathbf{KK}^{\mathsf T}\right) \nonumber \\ 
    &\mathrm{subject~to}~\left|\left|\left(\left|\left|\mathbf{k}_1\right|\right|_2,\cdots,\left|\left|\mathbf{k}_n\right|\right|_2\right)\right|\right|_0\leq p,~\mathbf{KU=I}, 
    \label{eq:objorgGL0f}
\end{align}
By taking the transpose of this formulation ($\mathbf{K}^{\mathsf T}\Rightarrow \mathbf{X}$, $\mathbf{U}^{\mathsf T}\Rightarrow \mathbf{A}$), we can obtain the following objective function. Here, the group $L_1$-norm penalty is adopted as an example.
\begin{align}
    \mathrm{minimize}~\mathrm{tr}\left(\mathbf{X}^{\mathsf T}\mathbf{X}\right)+\lambda \sum_{i=1}^n \left|\left|{\mathbf x}_i\right|\right|_2~~~~\mathrm{subject~to}~\mathbf{AX=I}
    \label{eq:obj}
\end{align}

\subsubsection{Alternating Direction Method of Multipliers} %soft, hard, hard number(検討)
Equation~\ref{eq:obj} is following form,
\begin{align}
    \mathrm{minimize}~~g(\mathbf{X})+h(\mathbf{Z})~~~~\mathrm{subject}~\mathrm{to}~~\mathbf{Z=GX},
\end{align}
and this optimization problem is efficiently solved by ADMM as follows:
\begin{align}
    \mathbf{X}^{(n+1)}&=\mathrm{arg}~\mathrm{min}~g\left(\mathbf{X^{(n)}}\right)+\frac{1}{2\gamma}\left|\left|\mathbf{Z}^{(n)}-\mathbf{GX^{(n)}}-\mathbf{Y}^{(n)}\right|\right|^2, \label{eq:argmin} \\
    \mathbf{Z}^{(n+1)}&=\mathrm{prox}_{\gamma h}~\left(\mathbf{GX}^{(n+1)}+\mathbf{Y}^{(n)}\right), \\
    \mathbf{Y}^{(n+1)}&=\mathbf{Y}^{(n)}+\mathbf{GX}^{(n+1)}-\mathbf{Z}^{(n+1)}.
\end{align}
In the present formulation with group $L_1$-norm penalty, $g(\mathbf{X})$, $h(\mathbf{Z})$, $\mathbf{Z}$, and $\mathbf{G}$ correspond to
\begin{align}
    g\left(\mathbf{X}\right)=\mathrm{tr}\left(\mathbf{X}^{\mathsf T}\mathbf{X}\right),&~h\left(\mathbf{Z}\right)=\lambda \sum_i \left|\left|\mathbf{x}_i\right|\right|_2, \\
    \mathbf{Z}=\left[
    \begin{array}{cc}
         \mathbf{Z_1}\\ \mathbf{Z_2}
    \end{array}\right]
    , {\rm and} &~\mathbf{G}=\left[
    \begin{array}{cc}
        \mathbf{I}\\ \mathbf{A}
    \end{array}\right],
\end{align}
respectively. In the present formulation, the function $g({\mathbf X})$ is differentiable, and thus, the solution of Eq.~\ref{eq:argmin} can be obtained as follows:
\begin{align}
    \frac{\mathrm d}{{\mathrm d}{\mathbf X}}\left\{\mathrm{tr}\left(\mathbf{X}^{\mathsf T}\mathbf{X}\right)+\frac{1}{2\gamma}\left|\left|\mathbf{Z}^{(n)}-\mathbf{GX}-\mathbf{Y}^{(n)}\right|\right|^2\right\}&=0, \nonumber \\
    2{\mathbf X}+\frac{1}{\gamma}{\mathbf G}^{\mathsf T}{\mathbf G}{\mathbf X}+\frac{1}{\gamma}{\mathbf G^{\mathsf T}\left({\mathbf Y}-{\mathbf Z}\right)}&=0.
\end{align}
Solving the above equation for $\mathbf X$, we can obtain the solution of Eq.~\ref{eq:argmin}.
\begin{align}
    {\mathbf X}&=\left(2{\mathbf I}+\frac{1}{\gamma}{\mathbf G}^{\mathsf T}{\mathbf G}\right)^{-1}\frac{1}{\gamma}{\mathbf G}^{\mathsf T}\left({\mathbf Z}-{\mathbf Y}\right) \nonumber \\ 
    &= {\scriptstyle \left\{\left(2+\frac{1}{\gamma}\right){\mathbf I}+\frac{1}{\gamma}\left({\mathbf A}^{\mathsf T}{\mathbf A}\right)\right\}^{-1}\frac{1}{\gamma}\left\{\left({\mathbf Z_1}-{\mathbf Y_1}\right)+{\mathbf A}^{\mathsf T}\left({\mathbf Z_1}-{\mathbf Y_1}\right)\right\} }
\end{align}
Here, the computational cost of the least-squares solution of Eq.~\ref{eq:argmin} can be reduced by adopting the inverse matrix lemma as shown below when the latent variables are sparse:
\begin{align}
    &\left\{\left(2+\frac{1}{\gamma}\right){\mathbf I}+\frac{1}{\gamma}\left({\mathbf A}^{\mathsf T}{\mathbf A}\right)\right\}^{-1} \nonumber \\
    &={\scriptstyle \frac{1}{2+\left(\frac{1}{\gamma}\right)}-\frac{1}{2+\left(\frac{1}{\gamma}\right)}{\mathbf A}^{\mathsf T}\left\{{\mathbf I}+\frac{1}{\gamma}{\mathbf A}\frac{1}{\left(2+\frac{1}{\gamma}\right)}{\mathbf A}^{\mathsf T}\right\}^{-1}\frac{1}{\gamma}{\mathbf A}\frac{1}{\left(2+\frac{1}{\gamma}\right)}.}
\end{align}
%where the inverse matrix lemma is
%\begin{align}
%    \left({\mathbf A}+{\mathbf B}{\mathbf C}\right)^{-1}={\mathbf A}^{-1}-{\mathbf A}^{-1}{\mathbf B}\left({\mathbf I}+{\mathbf C}{\mathbf A}^{-1}{\mathbf B}\right)^{-1}{\mathbf C}{\mathbf A}^{-1}.
%\end{align}

The sparsity-promoting term, the second term of Eq.~\ref{eq:obj}, was computed in three kinds of ways, which are the block soft thresholding, block hard thresholding, and $L_0$-norm constrained block hard thresholding operators. Here, $L_0$-norm constrained block hard thresholding is the hard thresholding operator that can determine the $L_0$ norm of the solution beforehand. In the present formulation, the group $L_0$ norm of the solution matrix corresponds to the number of selected sensors. The sparsity ($L_0$ norm) of the solution is related to the sparsity parameter $\lambda$ for conventional soft and hard thresholding operators, and $L_0$ norm cannot determine beforehand. Hence, the $L_0$-norm constrained operator is convenient to apply the sensor selection problem.

The proximal operator of the group $L_1$-norm penalty used in Eq.~\ref{eq:objorgGL1} corresponds to the block soft thresholding as shown below:
\begin{align}
   S_{\gamma\lambda}\left({\mathbf v}_i\right):=
    \left\{
    \begin{array}{cc}
        {\mathbf v}_i-\gamma\lambda\frac{{\mathbf v}_i}{||{\mathbf v}_i||_2},&~~~||{\mathbf v}_i||_2~\geq\gamma\lambda, \\
        0,&~~~||{\mathbf v}_i||_2~<\gamma\lambda. \\
    \end{array}
    \right.
\end{align}
The block hard thresholding is adopted as a proximal operator of the group $L_0$-norm penalty used in Eq.~\ref{eq:objorgGL0}.
\begin{align}
    H_{\gamma\lambda}\left({\mathbf v}_i\right):=
    \left\{
    \begin{array}{cc}
        {\mathbf v}_i,&~~~||{\mathbf v}_i||_2~\geq\gamma\lambda \\
        0,&~~~||{\mathbf v}_i||_2~<\gamma\lambda \\
     \end{array}
     \right.
\end{align}
The proximal operator of the $L_0$-norm constrained penalty used in Eq.~\ref{eq:objorgGL0f} is also the block hard thresholding but the threshold is $||{\mathbf v}_p||_2$, which is the $p$-th largest value of $||{\mathbf v}_i||_2$.
\begin{align}
    H_p\left({\mathbf v_i}\right):=
    \left\{
    \begin{array}{cc}
        {\mathbf v}_i,&~~~||{\mathbf v}_i||_2~\geq\||{\mathbf v}_p||_2 \\
        0,&~~~||{\mathbf v}_i||_2~<||{\mathbf v}_p||_2 \\
     \end{array}
     \right.
\end{align}
A similar hard thresholding operator can be seen in \cite{blumensath2008iterative}.

\subsubsection{Polishing Step}
Although the gain matrix $\mathbf{K}$ obtained through ADMM can be directly used, its performance is considered to be degraded to some extent by the sparsity promoting terms. Therefore, the sparsity pattern of the sensor selection is only utilized and the better gain is recalculated by using Eq.~\ref{eq:LS_estimation} rather than by using the gain matrix $\mathbf{K}$ directly. 
The sensor location matrix $\mathbf{H}$, which indicates the sparse pattern of sensors, is constructed based on the obtained gain matrix $\mathbf{K}$ obtained by optimization using ADMM. The corresponding entry component of $\mathbf{H}$ is set to be unity (activate the corresponding sensor) when the $L_2$ norm of the column vector of $\mathbf{K}$ is nonzero. In the present study, the sensor is activated when the $L_2$ norm of the column vector of $\mathbf{K}$ is greater than $10^{-4}$. The latent state variable $\tilde{\mathbf{z}}$ is estimated using Eq.~\ref{eq:LS_estimation}.

\section{Results and Discussion}
The performance of the proposed method was evaluated by applying to random sensor problems and compared with previously proposed methods. The random sensor-candidate matrix $\mathbf{U}\in \mathbb{R}^{n{\times}r}$ were generated by the Gaussian distribution of $\mathcal{N}(0,1)$. The convex relaxation, the greedy method, and the proposed method, which is based on the ADMM algorithm, were applied. The objective function was the minimization of the inverse of FIM in all computations. The computation for each condition was conducted 100 times with different sensor-candidate matrices and was averaging.

\subsection{Effect of Sparsity Parameter on Number of Selected Sensors}
\begin{figure}[]
\centering
\includegraphics[scale=0.7]{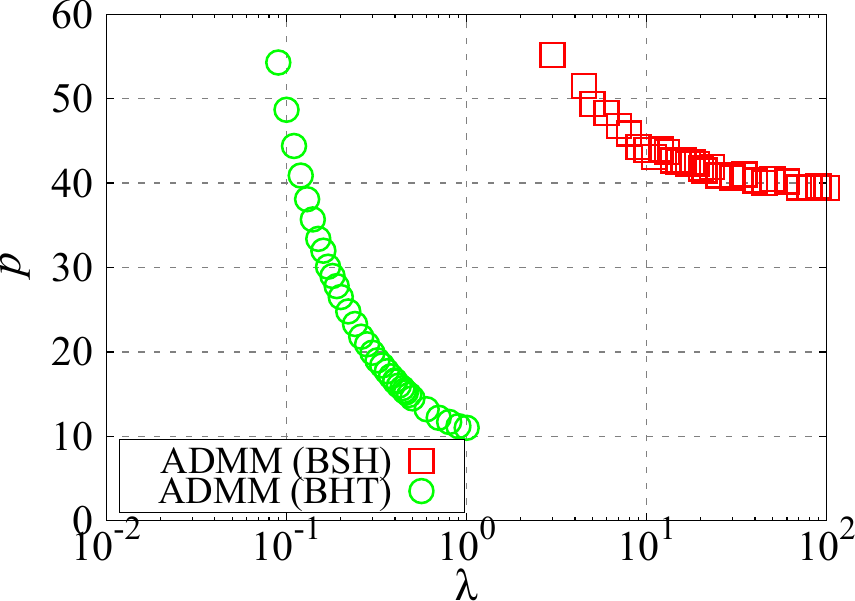}
\caption{Relationship between sparsity parameter and the number of selected sensors ($n=1000, r=10$).}
\label{fig:lambda-p}
\end{figure}

The number of sensors selected by the proposed method is related to $\lambda$ when the block soft thresholding and the block hard thresholding are used as a sparsity-promoting term, and the number of sensors cannot be determined beforehand. Figure~\ref{fig:lambda-p} shows the relationship between the sparsity parameter and the number of selected sensors when $\mathbf{U}\in\mathbb{R}^{1000{\times}10}$. 

The number of selected sensors decreases as the sparsity parameter increases. The block hard thresholding can determine the sparse sensor location for $p\geq 10$ in the preset problem setting. On the other hand, the block soft thresholding could not shrinking the solution compared with the block hard thresholding and could not determine the location of sparse sensors (the number of sensors becomes zero) at $p\lesssim 40$.

\subsection{Value of A-optimality Criterion}
The performance of the proposed methods is compared with the greedy method \cite{nakai2020effect} and the convex relaxation (see Appendix) as shown in Figure~\ref{fig:p-trinv}. Both the greedy method and the convex relaxation is  based on the A-optimal design of experiment. The trace of the inverse of FIM which is the objective function of the present method was computed. The sensor-candidate matrix was $\mathbf{U}\in\mathbb{R}^{1000{\times}10}$. The vertical axis is the normalized trace of the inverse of FIM which is normalized by the result of the greedy method. The computations using the ADMM-based method were conducted with the block soft thresholding (BST), the block hard thresholding (BHT), and the $L_0$-norm constrained block hard thresholding ($L_0$BHT). The step size $\gamma$ for ADMM was set to be $\gamma=0.4$. When the BHT and $L_0$BHT are used, convergence is not guaranteed depending on the value of $\gamma$ because the problem becomes a nonconvex problem. Therefore, $\gamma$ was gradually decreased in the computation using BHT and $L_0$BHT. The value of $\gamma$ was multiplied by 0.99 every 200 iterations.  It should be noted that the performance of ADMM with BHT and $L_0$BHT were better for larger $\gamma$ within the convergence range for most cases.

The performance of ADMM with BST is inferior (larger trace of the inverse of FIM) to any other compared methods. On the other hand, ADMM with BHT and $L_0$BHT can minimize the trace of the inverse of FIM in the most range of $p$ which is investigated in the present study, and these methods have almost the same performance and trend. The value of normalized trace of the inverse of FIM is relatively higher at larger $p$ but still have the best performance. The performances of ADMM with BHT and $L_0$BHT are improved with decreasing of $p$. The value of normalized trace of the inverse of FIM takes the minimum at approximately $p=15$, and it rapidly increases as the number of sensors further decreases. The performance of the convex relaxation also gets worsens at approximately $p=r$.

\begin{figure}[]
\centering
\includegraphics[scale=0.7]{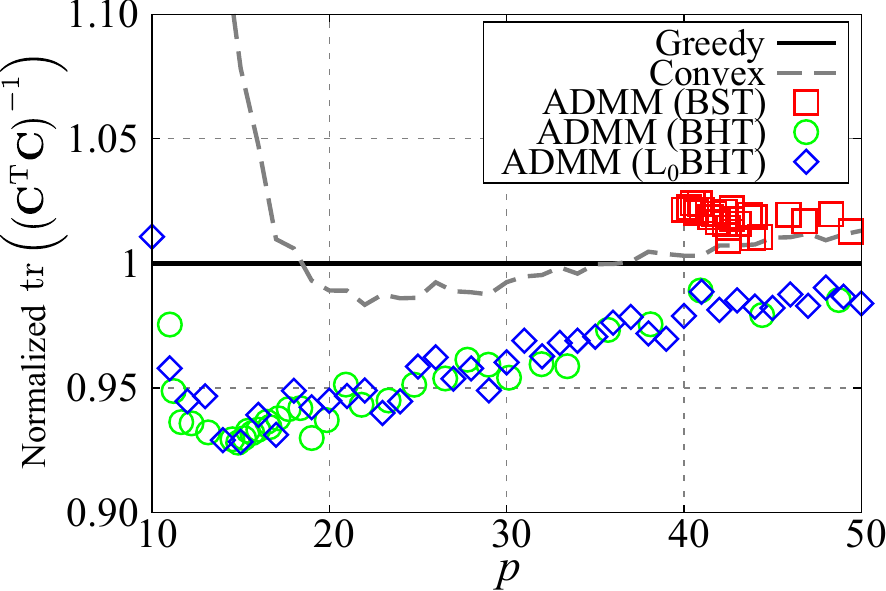}
\caption{Normalized trace of the inverse of FIM ($n=1000, r=10$).}
\label{fig:p-trinv}
\end{figure}

\subsection{Computational Cost}
The computational complexity of the previous and proposed methods are shown in Table~\ref{tab:complexity}. Considered problem in the present study is $r\leq p\ll n$. The computational complexity of the ADMM-based method is the first order of $n$ and square orders of $r$, and the convex relaxation is cubic orders of $n$. Therefore, the computational cost of the greedy method and the proposed method is quite smaller than that of the convex relaxation. The computational cost of the proposed method does not depend on $p$. However, the present method requires iteration, and the total computational cost of the present method is larger than that of the greedy method.

\begin{table}[]
\centering
\caption{Comparison of computational complexity.}
\begin{tabular}{@{}ll@{}}
\toprule
Method & Computational complexity \\ \midrule
Greedy method & ${\mathcal O}(pnr^2)$ \\
Convex relaxation & ${\mathcal O}(n^3)$ per iteration \\
ADMM-based method & ${\mathcal O}(nr^2)$ per iteration \\ \bottomrule
\end{tabular}
\label{tab:complexity}
\end{table}

The computational time was measured in the random sensor problem. The sensor-candidate matrix was $\mathbf{U}\in\mathbb{R}^{n{\times}10}$ and the number of sensors $p$ was fixed at $p=20$. The computational time $t_{\rm ave}$ is the average value of 100 times computations with different sensor-candidate matrices. The step size $\gamma$ of ADMM was fixed at $\gamma=0.2$. The numerical test was conducted on a desktop PC (CPU: Intel(R) Core(TM) i7-6800K 3.40 GHz; RAM 128 GB).

The computational time of ADMM with $L_0$BHT is larger than any other method for $n=10^2$. The computational time of the convex relaxation rapidly increases as $n$ increases. The greedy method and the proposed method are quite faster than the convex relaxation for a larger scale problem. The difference in the computational time of the greedy method and the proposed method becomes large as $n$ increases because of the increase in the number of iterations. The computational time of the proposed method is approximately 10 times larger than that of the greedy method at $n=10^5$ but the present method is still feasible in this range. Also, the performance of the present method is better than that of the greedy method in terms of the A-optimality criterion in the most range of $p\leq50$ in the random sensor problem with $\mathbf{U}\in\mathbb{R}^{1000{\times}10}$ as discussed in the previous section.

\begin{figure}[]
\centering
\includegraphics[scale=0.7]{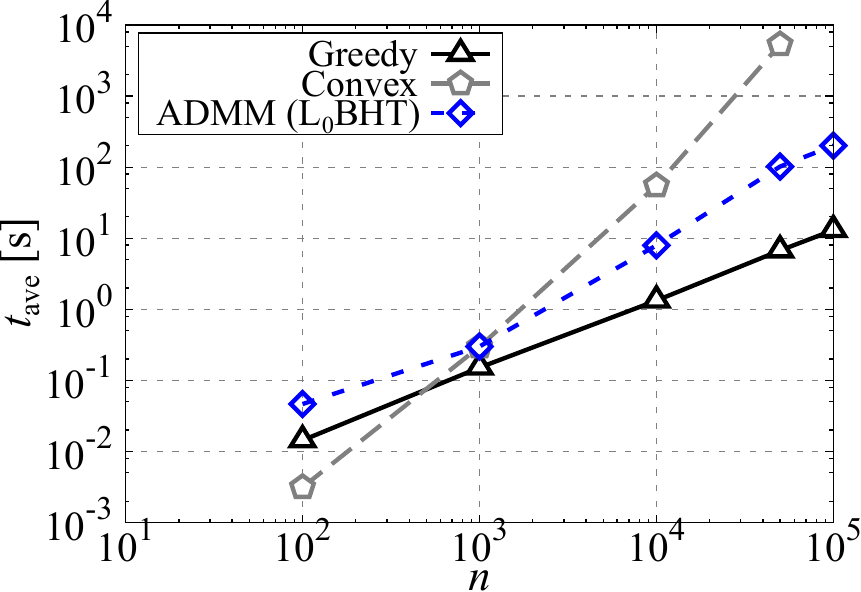}
\caption{Averaged computational time for different size datasets ($p=20, r=10$).}
\label{fig:n-time}
\end{figure}

\section{Application to Data-driven Sensor Selection}
The proposed method was applied to the data-driven sparse-sensor-selection problem. A data set adopted is the NOAA OISST (NOAA-SST) V2 mean sea surface temperature set, which comprises weekly global sea surface temperature measurements in the years between 1990 and 2000 \cite{noaa}. The dimensional reduction was conducted by the truncated singular value decomposition. A rank-$r$ approximation was obtained by keeping the leading $r$ singular values and vectors. The data matrix $\mathbf{X}\in\mathbb{R}^{n\times m}$, which consists of $m$ snapshots with spatial dimension $n$, is decomposed into the left singular matrix $\mathbf{U}\in\mathbb{R}^{n\times m}$, which shows the spatial POD modes, the diagonal matrix of singular values $\mathbf{S}\in\mathbb{R}^{m\times m}$, and the right singular matrix $\mathbf{V}\in\mathbb{R}^{m\times m}$, which shows temporal POD, and the rank-$r$ reduced-order modeling of a data matrix is given as $\mathbf{X}\approx\mathbf{U}_{1:r}\mathbf{S}_{1:r}\mathbf{V}_{1:r}^{\mathsf T}$. The measurement matrix $\mathbf{C}$ is $\mathbf{HU}_{1:r}$ and the latent state variable is the POD mode amplitude $\mathbf{Z}=\mathbf{S}_{1:r}\mathbf{V}_{1:r}^{\mathsf T}$. Here, the dimension of the data matrix was reduced with $r=10$. The employed data set consists of 520 snapshots on a $360\times180$ spatial grid. The temperature data of ten years was split into every two-year segment, and the five-fold cross-validation was conducted. The training data and the test data, $\mathbf{X_{\rm train}}$ and $\mathbf{X_{\rm test}}$, were 80\% and 20\% of the original data, respectively. The trace of the inverse of FIM and the reconstruction error were ensemble averaged.

\begin{figure}[]
\centering
\includegraphics[scale=0.6]{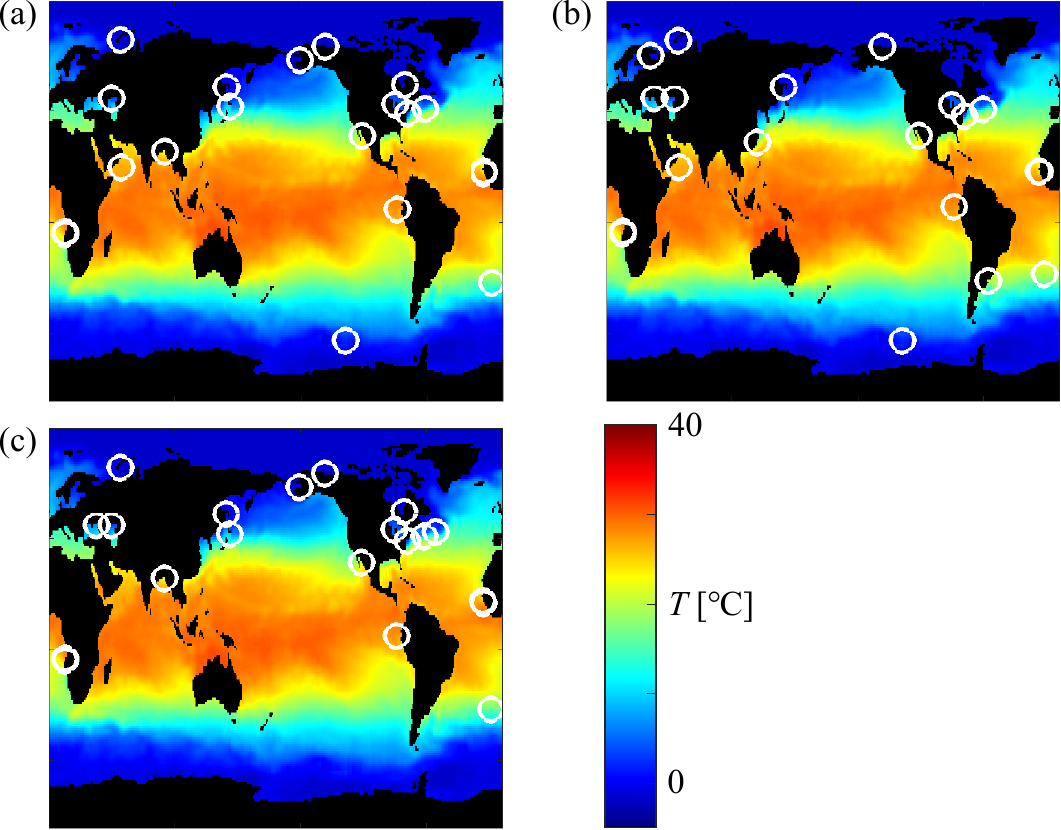}
\caption{Locations of the selected sensors ($p=20$). (a) convex relaxation; (b) greedy method; (c) ADMM (L$_0$BHT). The training data is between 1990 and 1998, and the test data was between 1998 and 2000.}
\label{fig:noaa-sensor}
\end{figure}

Locations of the selected sensors are displayed in Figure~\ref{fig:noaa-sensor}. The sensors were selected by the convex relaxation, the greedy method, and the ADMM-based method with L$_0$BHT. The objective function of the convex relaxation and the greedy method is the A-optimality criterion. Several sensors are placed in almost or exactly the same locations by each method, and several sensors are located in different places. 

\begin{figure}[]
\centering
\includegraphics[scale=0.7]{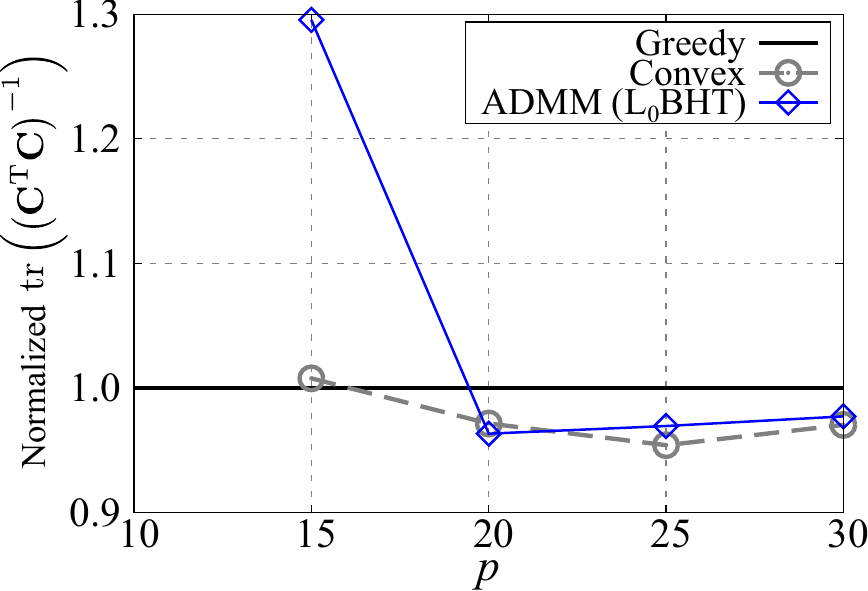}
\caption{The effect of the number of sensors on the trace of the inverse of FIM. The reference case is the result obtained by the greedy method.}
\label{fig:noaa-trinv}
\end{figure}

Figure~\ref{fig:noaa-trinv} illustrates the relationship between the number of sensors and the trace of the inverse of FIM. The trace of the inverse of FIM is normalized by the value obtained by the greedy method. The convex relaxation and the proposed method can minimize the trace of the inverse of FIM at $p\geq 20$ compared to the greedy method. The trace of the inverse of FIM obtained by the convex relaxation and the proposed method becomes larger than that of the greedy method at $p=15$, particularly, the increment of the proposed method is quite larger. 

\begin{figure}[]
\centering
\includegraphics[scale=0.7]{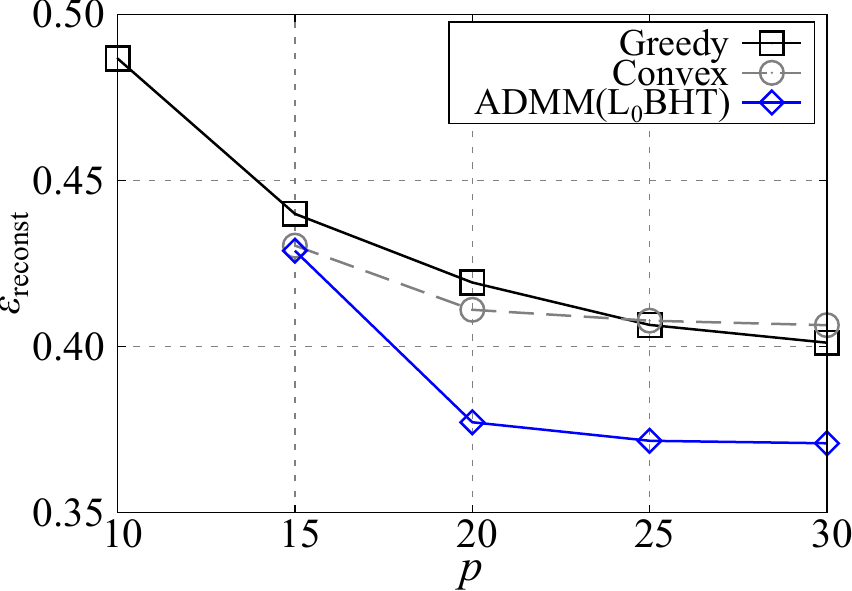}
\caption{The effect of the number of sensors on the reconstruction error. The reference case is the result obtained by the greedy method.}
\label{fig:noaa-reconsterr}
\end{figure}

\begin{table}[]
\centering
\caption{Comparison of computational time ($p=20$).} %The training data was between 1990 and 1998, and the test data was between 1998 and 2000.}
\begin{tabular}{@{}ll@{}}
\toprule
Method & Computational time \\ \midrule
Greedy method & 5.5~s \\
Convex relaxation & 77430.2~s \\
ADMM-based method & 320.4~s \\ \bottomrule
\end{tabular}
\label{tab:noaa-time}
\end{table}

%再構築誤差
Figure~\ref{fig:noaa-reconsterr} shows the effect of the number of sensors on the reconstruction error. The definition of the reconstruction error is following equation.
\begin{align}
\epsilon_{\rm reconst}=\frac{1}{m}\sum_{i=1}^m\frac{||\mathbf{\tilde{x}}_i-\mathbf{x}_i||_2}{||\mathbf{x}_i||_2}
\end{align}
Here, $\mathbf{x}_i$ is the column vector of the test data matrix $\mathbf{X}$, and $\mathbf{\tilde{x}}$ is computed as follows:
\begin{align}
    \mathbf{\tilde{x}}=\mathbf{U}_{1:r}\mathbf{\tilde{z}}
\end{align}
The amplitude of modes, the latent state variable, is estimated by Eq.~\ref{eq:LS_estimation}. The reconstruction error decreases as the number of sensors increases. The results obtained by the greedy method and the convex relaxation are similar and the proposed method is smaller, particularly at $p\geq 20$. The increment in the reconstruction error from $p=20$ to 15 is quite large. This reflects that the trace of the inverse of FIM for the proposed method gets worse from $p=20$ to $p=15$ as discussed in Figure~\ref{fig:noaa-trinv}. However, the reconstruction error of the proposed method is still smaller at $p=15$.

\section{Conclusions}
The present paper proposed the sparse sensor selection method based on the proximal splitting algorithms and the A-optimal design of experiment. The proposed method determines the sensor location by minimizing the trace of the inverse of FIM with the constraint condition. The three kinds of sparsity-promoting terms, which are the block soft thresholding, block hard thresholding, and $L_0$-constrained block hard thresholding, were adopted.

The performance of the proposed method was compared with the greedy method and the convex relaxation based on the A-optimal desing of experiment in the random sensor problem. The result of the numerical test showed that proposed method with block hard thresholding and $L_0$-norm constrained block hard thresholding show the best performance in the most range of the oversampling region ($10\leq p\leq 50$). Particularly, the performance of the proposed method becomes better in the range of $p$ which is smaller but not close to $p=r$.

The effect of the size of the sensor-candidate matrix on the computational time was investigated. In the small scale problem (e.g. $n=10^2$), the computational time of the proposed method is larger than that of the greedy method and the convex relaxation. The computational complexity of the proposed method is ${\mathcal O}\left(nr^2\right)$ per iteration, and thus, the proposed method is advantageous for a large-scale problem in terms of the computational cost compared with the convex relaxation. %The computational cost of the present method does not depend on the number of sensors but it is an iterative method. The present method requires longer computational time than the greedy method. However, 
The computational cost at a large scale problem (e.g., $n=10^5$) is feasible.

The proposed method was applied to a large-scale practical dataset, NOAA OISST (NOAA-SST) V2 weekly mean sea surface temperature. The proposed method shows better performance compared to the greedy method and the convex relaxation in terms of the reconstruction error. Although the greedy method is still faster than the proposed method, the computational time of the proposed method is quite shorter than that of the convex relaxation. However, the proposed method is an iterative method, and thus, solutions would be faster acquired by adjusting the threshold for convergence judgment and step size of ADMM.

\section*{Acknowledgement}
This work was supported by JST CREST Grant Number JPMJCR1763.

\section*{Appendix: convex relaxation method based on the A-optimality criterion}
We straightforwardly extended the convex relaxation method proposed by Joshi and Boyd \cite{joshi2009sensor}, as an existing algorithm compared. The objective function of the original method is the maximization of the logarithm of the  determinant of FIM.

The approximate relaxed sensor selection problem with minimization of trace of the inverse of FIM can be written as follows:
\begin{align}
    {\rm minimize}~~~\Psi\left(z\right)&={\rm tr}\left(\left(\sum_{i=1}^m \mathbf{z}_i\mathbf{u}_i\mathbf{u}_i^{\mathsf T}\right)^{-1}\right) \nonumber \\
    &~~~~-\kappa \sum_{i=1}^m\left({\rm log}\left(z_i\right)+{\rm log}\left(1-z_i\right)\right) \nonumber \\
    &{\rm subject~to}~~~{\mathbf 1}^{\mathsf T}\mathbf{z}=p.
\label{eq:trinvonv}
\end{align}
The variable $\mathbf{z}\in\mathbb{R}^m$ is the solution vector with implicit constraints that $z_i\in(0,1)$, which corresponds to weights for sensor candidates. A positive parameter $\kappa$ that controls the quality of approximation. A set of $m$ potential measurements is characterized by $\mathbf{u}_1,\cdots,\mathbf{u}_m\in\mathbb{R}^n$. The objective function $\Psi$ is concave and smooth. Therefore, Eq.~\ref{eq:trinvonv} can be efficiently solved by Newton's method same as the formulation of Joshi and Boid \cite{joshi2009sensor}.

The derivatives of $\Psi$ is given below. The gradient is given by 
\begin{align}
    \left(\nabla\Psi\right)_i=-\mathbf{u}_i^{\mathsf T}W^{-2}\mathbf{u}_i-\left(\frac{\kappa}{z_i}-\frac{\kappa}{1-z_i}\right),~~i=1,\cdots, m
\end{align}
where
\begin{align*}
    \mathbf{W}=\sum_{i=1}^mz_i\mathbf{u}_i\mathbf{u}_i^{\mathsf T}.
\end{align*}
The Hessian $\nabla^2\Psi$ is given by
\begin{align}
    &\nabla^2\Psi=2\left(\mathbf{AW}^{-2}\mathbf{A}^{\mathsf T}\right)\circ\left(\mathbf{AW}^{-1}\mathbf{A}^{\mathsf T}\right) \nonumber \\
    &~~+\kappa~{\rm diag}~\left(\frac{1}{z_1^2}+\frac{1}{\left(1-z_1\right)^2},\cdots,\frac{1}{z_m^2}+\frac{1}{\left(1-z_m\right)^2}\right)
\end{align}
where $\circ$ denotes the Hadamard product and $\mathbf{U}_{1:m}$ is the measurement matrix
\begin{align*}
    \mathbf{U}_{1:m}=\left[
    \begin{array}{c}
    \mathbf{u}_1^{\mathsf T} \\
    \vdots \\
    \mathbf{u}_m^{\mathsf T} \\
    \end{array}
    \right].
\end{align*}

We stop the computation when the Newton decrement is less than $10^{-4}$. The total number of steps for Newton's method in this formulation required approximately 20 for random sensor problem, required approximately 300 for NOAA-SST problem. After getting the converged solution, sensors with $p$-largest $z_i$ values are selected.

\bibliographystyle{IEEEtran}
\bibliography{xaerolab}

\begin{IEEEbiography}
[{\includegraphics[width=1in,height=1.25in,clip,keepaspectratio]{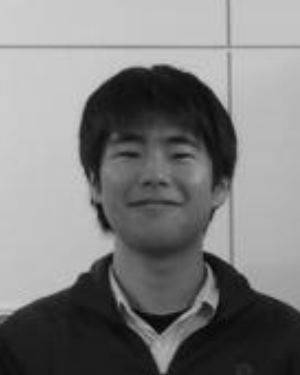}}]{Takayuki Nagata} received the B.S. degree in mechanical and aerospace engineering from Tokai University, Japan in 2015, and the Ph.D. degree in aerospace engineering from Tohoku University, Japan, in 2020. From 2018-2020, he was a Research Fellow of Japan Society for the Promotion of Science (JSPS) at Tohoku University, Japan. Since 2020, he has been a postdoctoral researcher at Tohoku University, Japan. 
\end{IEEEbiography}

\begin{IEEEbiography}
[{\includegraphics[width=1in,height=1.25in,clip,keepaspectratio]{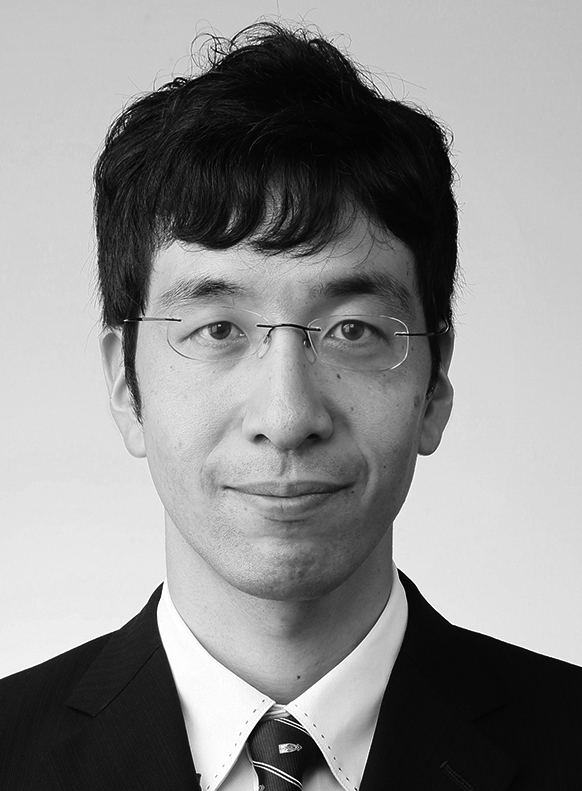}}]{Taku Nonomura} received the B.S. degree in mechanical and aerospace engineering from Nagoya University, Nagoya, Japan in 2003, and the Ph. D. degree in aerospace engineering from the University of Tokyo, Tokyo, Japan in 2008. He is currently an Associate Professor of Aerospace Engineering at Tohoku University, Sendai.
\end{IEEEbiography}

\begin{IEEEbiography}
[{\includegraphics[width=1in,height=1.25in,clip,keepaspectratio]{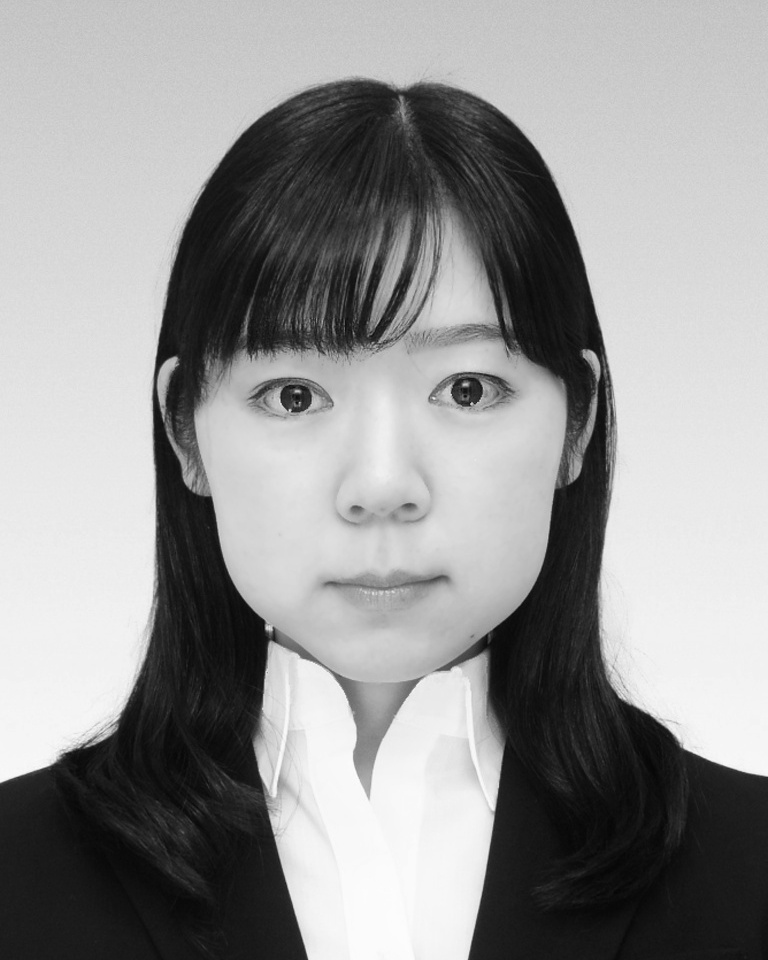}}]{Kumi Nakai} received the Ph.D. degree in mechanical systems engineering from Tokyo University of Agriculture and Technology, Japan, in 2020. From 2017-2020, she was a Research Fellow of Japan Society for the Promotion of Science (JSPS) at Tokyo University of Agriculture and Technology, Japan. Since 2020, she has been a postdoctoral researcher at Tohoku University, Japan. Her research interests include data-driven science, discharge plasma dynamics, fluid dynamics, and flow control utilizing atmospheric pressure plasma. 
\end{IEEEbiography}

\begin{IEEEbiography}
[{\includegraphics[width=1in,height=1.25in,clip,keepaspectratio]{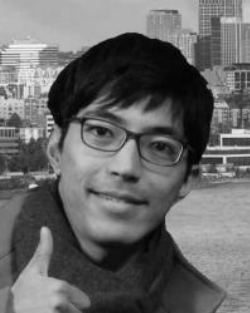}}]{Keigo Yamada} received the B.S. degree in physics from Tohoku University, Sendai, Japan, in 2019. He is a M.S. student of engineering at Tohoku University.
\end{IEEEbiography}

\begin{IEEEbiography}
[{\includegraphics[width=1in,height=1.25in,clip,keepaspectratio]{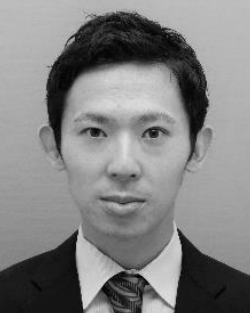}}]{Yuji Saito} received the B.S. degree in mechanical engineering, and the Ph.D. degree in mechanical space engineering from Hokkaido University, Sapporo, Japan in 2018. He is an Assistant Professor of Aerospace Engineering at Tohoku University, Sendai, Japan.
\end{IEEEbiography}

\begin{IEEEbiography}
[{\includegraphics[width=1in,height=1.25in,clip,keepaspectratio]{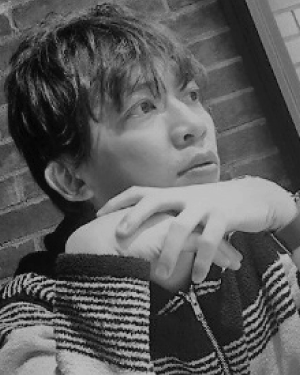}}]{Shunsuke Ono} received the B.E. degree in computer science, in 2010, and the M.E. and Ph.D. degrees in communications and computer engineering, in 2012 and 2014, respectively, from the Tokyo Institute of Technology. From April 2012 to September 2014, he was a Research Fellow (DC1) of the Japan Society for the Promotion of Science (JSPS). He is currently an Associate Professor with the Department of Computer Science, School of Computing, Tokyo Institute of Technology. Since 2016.
\end{IEEEbiography}

\end{document}